\documentclass[prd,nofootinbib,english]{revtex4}

\usepackage{graphicx,float}
\usepackage{amsmath,amssymb,amsfonts}
\usepackage{epsfig,color}
\usepackage[thinlines]{easytable}
\usepackage{pdfpages}
\usepackage{array}
\usepackage{cancel}
\usepackage{mathtools}
\usepackage{accents}
\usepackage{subfigure}
\usepackage{enumitem}
\usepackage[dvipsnames]{xcolor}
\usepackage{refstyle}
 \usepackage{hyperref}
\hypersetup{
    colorlinks=true,
    linkcolor=blue,
    filecolor=magenta,      
   citecolor=blue
}

\newcommand{\udt}[3]{#1^{#2}_{\phantom{#2}#3}}
\newcommand{\udut}[4]{#1^{#2\phantom{#3}#4}_{\phantom{#2}#3\phantom{#4}}}

\newcommand{\dut}[3]{#1_{#2}^{\phantom{#2}#3}}
\newcommand{\dudt}[4]{#1_{#2\phantom{#3}#4}^{\phantom{#2}#3}}
\newcommand{\lc}[1]{\accentset{\circ}{#1}}

\begin{document}

\title{Generalised Proca Theories in Teleparallel Gravity}
\author{Gianbattista-Piero Nicosia}
\email{gianbattista-piero.nicosia.15@um.edu.mt}
\affiliation{Institute  for  Theoretical  Physics, University  of  Amsterdam,  1090  GL  Amsterdam,  The  Netherlands}
\affiliation{Institute of Space Sciences and Astronomy, University of Malta, Malta}

\author{Jackson Levi Said}
\email{jackson.said@um.edu.mt}
\affiliation{Institute of Space Sciences and Astronomy, University of Malta, Malta}
\affiliation{Department of Physics, University of Malta, Malta}

\author{Viktor Gakis}
\email{vgakis@central.ntua.gr}
\affiliation{Institute of Space Sciences and Astronomy, University of Malta, Msida, Malta}
\affiliation{Department of Physics, National Technical University of Athens, Zografou Campus GR 157 73, Athens, Greece}

\begin{abstract}
Generalised Proca theories of gravity represent an interesting class of vector-tensor theories where only three propagating degrees of freedom are present. In this work, we propose a new teleparallel gravity analog to Proca theories where the generalised Proca framework is extended due to the lower order nature of torsion based gravity. We develop a new action contribution and explore the example of the Friedmann equations in this regime. We find that teleparallel Proca theories offer the possibility of a much larger class of models in which do have an impact on background cosmology.
\end{abstract}

\maketitle

\section{Introduction}
The last decades have shown numerous successes for $\Lambda$CDM where cold dark matter is responsible for the accurate reproduction of galactic dynamics \cite{Baudis:2016qwx} while cosmological scale physics is complemented by a cosmological constant, $\Lambda$, which together with inflation produces one possible cosmology \cite{Perenon:2015sla}. The gravitational foundations of $\Lambda$CDM is general relativity (GR) whereas the other contributors represent modifications to the standard model of particle physics \cite{weinberg2008cosmology}. However, a number of crucial observational results have recently started to show the limits of $\Lambda$CDM cosmology, which are in addition to well-known theoretical problems inherent in the theory \cite{Weinberg:1988cp,Clifton:2011jh}. The most prominent of these are the $H_0$ tension, where local observations and early Universe $\Lambda$CDM dependent observations predict differing values of the expansion rate at current times \cite{DiValentino:2020zio}, and the $f\sigma_8$ tension which is a growing tension in the growth of large scale structure in the Universe \cite{DiValentino:2020vvd}, which similarly suffers a difference in values between local observations and predictions from the early Universe. Another important observation is that of the possible cosmic birefringence in the Planck Collaboration data \cite{Minami:2020odp} which would constitute a clear division between GR and many of its proposed modifications.
\medskip

This motivates us to explore other possible ways to meet the observational challenge that the current state of the art poses. The Lovelock theorem \cite{Lovelock:1971yv} provides a clear pathway in which to probe the extra degrees of freedom that one would have to introduce to examine other possible gravitational theories as a foundation of modifying GR in some way. One of these potential avenues is that of teleparallel gravity (TG), where the curvature associated with the Levi-Civita connection is exchanged with the torsion produced by the teleparallel connection \cite{Weitzenbock1923,Aldrovandi:2013wha}. In this way, TG produces a novel framework in which to construct theories of gravity that does not depend on GR in its traditional form. One such theory is the teleparallel equivalent of general relativity (TEGR) which is dynamically equivalent to GR in terms of its field equations \cite{Krssak:2018ywd,Cai:2015emx} but stems from a wholly different action in which the Ricci scalar Lagrangian is curtailed in such a way to eliminate the total divergence terms \cite{Escamilla-Rivera:2019ulu,Franco:2020lxx}. Naturally, this will weaken the Lovelock theorem in TG \cite{Gonzalez:2015sha,Gonzalez:2019tky} by producing a much broader plethora of paths to constructing second order theories of gravity. TG also has a number of other attractive features such as its likeness to Yang-mills theories \cite{Aldrovandi:2013wha} giving it a more particle physics perspective, its possibility of giving a definition to the gravitational energy-momentum tensor \cite{Blixt:2018znp,Blixt:2019mkt}, and that it does not require the introduction of a Gibbons--Hawking--York boundary in order to produce a well defined Hamiltonian description, among several others. TG also has a number of observationally attractive features in many of its formulations \cite{Briffa:2020qli,LeviSaid:2020mbb,Finch:2018gkh,Bahamonde:2020bbc}. \medskip

Gravitational wave astronomy offers the possibility of a large number of interesting novel tests of gravity that may reveal more stringent constraints on possible fundamental physical theories of gravity. The most impactful of these has up to now been the measurement that gravitational waves propagation at the speed of light up to one part in $10^{15}$ \cite{TheLIGOScientific:2017qsa,Goldstein:2017mmi} which was the result of multimessenger observations of the merger of a binary neutron star system. Up to this point many of the proposals beyond $\Lambda$CDM had been grouped in the Horndeski gravity framework in the context of curvature based theories \cite{Clifton:2011jh}, i.e. most modifications to gravity had been shown to be dynamically equivalent to a model in this landscape of gravitational models. Horndeski gravity is the most general curvature based theory of gravity that contains a single scalar field but retains second order field equations \cite{Kobayashi:2011nu,Gleyzes:2013ooa,Koyama:2015vza}. However, the multimessenger signal constraint resulted in a severe limiting of the possible expressions of a Horndeski theory \cite{Creminelli:2017sry,Sakstein:2017xjx}. Recently, a TG analog of Horndeski gravity was proposed in Ref.\cite{Bahamonde:2019shr} where Bahamonde--Dialektopoulos--Levi Said (BDLS) constructed a larger class of second order theories in which only one scalar field was included but second order field equations were produced. This stems from the naturally lower order nature of TG. BDLS theory was then shown to produce an organic way to revive previously ruled out models by producing a much more general gravitational wave propagation equation in Ref.\cite{Bahamonde:2019ipm}. The framework was also shown to be largely compatible with solar system tests through its parameterized post-Newtonian formalism in  Ref.\cite{Bahamonde:2020cfv}. \medskip

On the other hand, the standard model of particle physics contains several abelian and non-abelian vector fields as fundamental fields representing gauge interactions \cite{weinberg_1995}. In this context, there is strong motivation to explore bosonic vector fields on the scale of cosmic evolution, for instance as the candidate for dark energy which produces the late time accelerated expansion \cite{Heisenberg:2017mzp,deRham:2020yet}. This may have important consequences for relating beyond $\Lambda$CDM theories to particle physics models, and to explaining certain phenomenology such as cosmic birefringence or cosmological principle tests \cite{Heisenberg:2018vsk}. These Proca theories can be generalised by the equivalence principle to generalised (gravitational) Proca theories \cite{Heisenberg:2014rta} where a single arbitrary vector field interacts with the gravitation sector while producing second order field equations that propagation three degrees of freedom. Vector-tensor theories can be shown to support isotropic solutions \cite{Heisenberg:2014rta,Allys2015} with a temporal vector field which also feature screening mechanisms. In fact, in Ref.\cite{Heisenberg:2018acv} the foundations of a possible unification of scalar and Proca fields into a single theory was explored. \medskip

TG offers a novel platform on which to construct theories of gravity beyond $\Lambda$CDM. In the context of Proca fields, TG will produce the same contributions as that of the Levi-Civita connection, due to its lower order nature, in addition to further contributions through extra coupling scalars between the vector field and the gravitational sector. As in standard Proca theories \cite{Heisenberg:2017mzp,Heisenberg:2014rta}, the action for a propagating massive spin-1 field carrying three degrees of freedom that is invariant under local Lorentz transformations and observes locality is given by
\begin{equation} \label{sp}
    \mathcal{S}_{P} = \int d^{4}x\sqrt{-g} \bigg[-\frac{1}{4}F_{\mu\nu}F^{\mu\nu} - \frac{1}{2}m^{2}A_{\mu}A^{\mu}\bigg]\,,
\end{equation}
where $\sqrt{-g} = {\rm det}(g_{\mu\nu})$ is the determinant of the metric tensor and $F_{\mu\nu} = \partial_{\mu}A_{\nu} - \partial_{\nu}A_{\mu}$. By demanding general coordinate invariance, we will explore the case of a TG Proca theory framework by first introducing the foundation of TG in section \ref{section:Teleparallel Gravity}. In section \ref{gen_proca_theores}, the standard picture of Proca theories is described for curvature based gravity, while the TG scenario is explored in section \ref{TG_proca_theory}. As an example, we explored the Friedmann equations for the theory that we develop here in section \ref{sec:Cosmological_background_in _teleparallel_Proca_Theories} where the vector conservation equation is also presented. Finally, we close with a summary and conclusion of the main results in section \ref{Conclusion}.

\section{Teleparallel Gravity\label{section:Teleparallel Gravity}}

GR is the most prominent theory of gravity and is based on the Levi-Civita connection $\mathring{\Gamma}^{\sigma}_{\mu\nu}$ (we use over-circles to denote quantities calculated with the Levi-Civita connection throughout) which is torsion-less and satisfies the metricity condition \cite{nakahara2003geometry,misner1973gravitation}.
There is another formulation of GR whereby shifting the geometrical content from curvature to torsion at the cost of changing the Levi-Civita connection to the teleparallel connection denoted as $\Gamma^{\sigma}_{\mu\nu}$. The teleparallel connection is defined as the unique connection that is torsion-ful, curvature-less and metrical \cite{ortin2004gravity,Aldrovandi:2013wha,Cai:2015emx,Krssak:2018ywd}. \medskip

GR and its modifications are based on curvature which is ultimately based on the metric $g_{\mu\nu}$ through the Levi-Civita connection and thus the fundamental variable. On the other hand in TG theories, torsion is defined by the tetrad field $\udt{e}{A}{\mu}$ and the spin connection $\udt{\omega}{A}{B\mu}$ rendering the metric a derived object \cite{Aldrovandi:2013wha}. We will use capital Latin letters to denote the  Minkowski indices while Greek indices will denote the spacetime manifold as usual. The tetrad links the spacetime manifold with a Minkowski manifold where the orthonormal frames (indicated by capital Latin letters) exist. In 
this way one can induce the spacetime metric by the Minkowski metric as
\begin{eqnarray}\label{metric_tetrad_eq}
    g_{\mu\nu}=\udt{e}{A}{\mu}\udt{e}{B}{\nu}\eta_{AB}\,,& &\eta_{AB}=\dut{E}{A}{\mu}\dut{E}{B}{\nu}g_{\mu\nu}\,,
\end{eqnarray}
where we denote $\dut{E}{A}{\mu}$ as the inverse tetrad. In addition, the tetrad and inverse tetrad are related through
\begin{eqnarray}
    \udt{e}{A}{\mu}\dut{E}{B}{\mu}=\delta^B_A\,,& &\udt{e}{A}{\mu}\dut{E}{A}{\nu}=\delta^{\nu}_{\mu}\,.
\end{eqnarray}

\noindent The TG connection can be explicitly defined using the tetrad and the spin connection as \cite{Weitzenbock1923}
\begin{equation}
    \Gamma^{\sigma}{}_{\nu\mu} \coloneqq  \dut{E}{A}{\sigma}\partial_{\mu}\udt{e}{A}{\nu} + \dut{E}{A}{\sigma}\udt{\omega}{A}{B\mu}\udt{e}{B}{\nu}\,,
\end{equation}
which is the most general linear affine connection that is flat and metrical \cite{Aldrovandi:2013wha,Hehl:1994ue}. The metricity condition is also reflected by $\omega_{AB\mu}=-\omega_{BA\mu}$ which also defines the class of Lorentz spin connections. The spin connection, in general is introduced to attain Lorentz covariance and for TG it is totally inertial and incorporates the effects of the local Lorentz transformations (LLTs) thus producing LLT invariant theories. One can always find a frame where the TG spin connection is vanishing as in the original formulation in Ref.\cite{Weitzenbock1923}, and this choice of frame is dubbed the Weitzenb\"{o}ck gauge (WG). The TG spin connection can be fully represented as $\udt{\omega}{A}{B\mu}=\udt{\Lambda}{A}{C}\partial_{\mu}\dut{\Lambda}{B}{C}$ \cite{Aldrovandi:2013wha}, where the full breadth of the LLTs (Lorentz boosts and rotations) are represented by $\udt{\Lambda}{A}{B}$. Hence there is no unique tetrad that induces a specific spacetime metric Eq.~(\ref{metric_tetrad_eq}). \medskip

Using the TG connection we can define the torsion tensor as 
\begin{equation}
    \udt{T}{\sigma}{\mu\nu} \coloneqq - 2\Gamma^{\sigma}_{\left[\mu\nu\right]}\,,
\end{equation}
where square brackets denote the usual antisymmetrization. This tensor field also acts as the field strength of TG \cite{Aldrovandi:2013wha}, which transforms covariantly under both diffeomorphisms and LLTs. We can consequently build the contorsion tensor as 
\begin{equation}\label{eq:contorsion}
    \udt{K}{\sigma}{\mu\nu} \coloneqq   \udt{\Gamma}{\sigma}{\mu\nu} - \udt{\mathring{\Gamma}}{\sigma}{\mu\nu} = \frac{1}{2}\left(\dudt{T}{\mu}{\sigma}{\nu} + \dudt{T}{\nu}{\sigma}{\mu} - \udt{T}{\sigma}{\mu\nu}\right)\,,
\end{equation}
that links the Levi-Civita connection and the TG connection. \medskip

This has an important role to play in relating TG with GR and its modifications, as will become apparent later on. Another core component of TG is the superpotential defined as \cite{Krssak:2018ywd}
\begin{equation}
    \dut{S}{\lambda}{\mu\nu} \coloneqq  \frac{1}{2}\left(\udt{K}{\mu\nu}{\lambda} + \dut{\delta}{\lambda}{\nu}T^{\mu} - \dut{\delta}{\lambda}{\mu}T^{\nu}\right)\,,
\end{equation}
where $T^{\nu}\coloneqq \udut{T}{\alpha}{\alpha}{\nu}=-\udt{T}{\alpha\nu}{\alpha}$. This tensor is related with the energy-momentum tensor for gravitation \cite{Aldrovandi:2004db} but the issue remains open \cite{Koivisto:2019jra}. Contracting the torsion tensor with the superpotential we define the torsion scalar as \cite{Cai:2015emx}
\begin{equation}\label{torsion_scalar_def}
    T \coloneqq  \dut{S}{\lambda}{\mu\nu}\udt{T}{\lambda}{\mu\nu}= \frac{1}{4}T_1 + \frac{1}{2}T_2 - T_3\,,
\end{equation}
where 
\begin{equation}
T_1 \coloneqq  T^{\mu\nu\rho}T_{\mu\nu\rho}\,, \quad
T_2 \coloneqq  T^{\mu\nu\rho}T_{\rho\nu\mu}\,, \quad
T_3 \coloneqq  T_{\rho}T^{\rho}\,.
\end{equation}
The torsion scalar plays the role of the Lagrangian density just like the Ricci scalar. The standard Ricci scalar $\lc{R}$ (computed with the Levi-Civita connection) can be expressed, by using  contorsion tensor Eq.(\ref{eq:contorsion}), as the torsion scalar plus some boundary term \cite{Hayashi:1979qx,Hehl:1976kj}
\begin{equation}\label{eq:Ricci_torsion_equiv}
    \lc{R}  = -T + B\,,
\end{equation}
where the boundary term is defined as
\begin{equation}\label{eq:Boundary_term_def}
    B\coloneqq 2\mathring{\nabla}_{\mu}\left(T^{\mu}\right)\,.
\end{equation}
It is a well known fact that Eq.(\ref{eq:Ricci_torsion_equiv}) is the evidence of dynamical equivalence between GR and TEGR and serves as a justification of using torsion as our primary geometrical object \cite{Aldrovandi:2013wha}. Following this paradigm shift, we define the TEGR action as
\begin{equation}\label{eq:TEGR_action}
    \mathcal{S}_{\text{TEGR}} \coloneqq  -\frac{1}{2\kappa^2}\int d^4 x\, eT + \int d^4 x\, e\mathcal{L}_m\,,
\end{equation}
where $e = \text{det}\left(\udt{e}{A}{\mu}\right)=\sqrt{-g}$ is the tetrad determinant, $\kappa^2=8\pi G$ and $\mathcal{L}_m$ is the regular matter Lagrangian. \medskip

The (linear) boundary term difference at the level of the Lagrangians plays no role regarding the field equations but this is not the case when it is introduced in a non-linear manner like in $f(T,B)$ gravity \cite{Bahamonde:2015zma,Capozziello:2018qcp,Bahamonde:2016grb,Paliathanasis:2017flf,Farrugia:2018gyz,Bahamonde:2016cul,Bahamonde:2016cul,Wright:2016ayu}. In fact, Eq.(\ref{eq:Ricci_torsion_equiv}) could be seen as a split of the Ricci scalar into a contribution of first order derivatives of the tetrad(Torsion scalar) and a contribution of only second order derivatives of the tetrad(Boundary term). \medskip

A common misconception in the realm of TG gravity is the use of the WG, where one chooses a tetrad field in such a way that the spin connection is trivialised. Upon trivializing, the spin connection the manifest LLT covariance is lost. This does not mean that the theory is Lorentz violating but rather that it is presented in a non-manifest LLT covariant form. We remind the reader that one can do exactly the same for the case of the Levi-Civita connection but only locally, i.e, choosing a particular coordinate system around a point where  $\mathring{\Gamma}^{\sigma}_{\mu\nu}\equiv0$ and thus trivialising the Levi-Civita connection at the cost of diffeomorphism covariance. Again this does not mean that the theory, truly, lost the diffeomorphism invariance but rather that we work on a specific coordinate system to simplify the analysis. One can always restore any type of the aforementioned covariances by re-introducing the connections by a proper transformation of the metric or tetrad. As a matter of fact, the flat spin connection of TG from the WG can be transformed into a non-trivial form through an arbitrary Lorentz transformation as
\begin{equation}\label{eq:tetrad_spin_con_trans}
    e^A{}_{\mu} \mapsto e'^A{}_{\mu} = \Lambda^A{}_Be^B{}_{\mu}\,,
\end{equation}
\begin{equation}
    \omega'^A{}_{B\mu} = \Lambda^A{}_C\partial_{\mu}(\Lambda^{-1})^C{}_B\,.
\end{equation}
Hence one can straight-forwardly restore Local Lorentz covariance. One last, quite potent, argument in favour of the use of the WG gauge in TG is that the spin connection is non-dynamical and thus it is just a gauge degrees of freedom, hence if one formulates the action by LLT invariant scalars then the spin connection can be safely trivialised. In the light of these arguments, we will work in the WG throughout this paper having implicitly trivialised the spin connection without loss of generality. \medskip

We close off this section by presenting the irreducible decomposition of the torsion tensor, under the action of the LLT, into a vector $v_{\mu}$, a pseudo-vector  $a_{\mu}$ and a pure tensor part $t_{\alpha\mu\nu}$ as
\begin{subequations}
\begin{eqnarray}\label{eq:Tor_ir_comp1}
v_{\mu} & \coloneqq  & T_{\mu},\\
a_{\mu} & \coloneqq  & \frac{1}{6}\epsilon_{\mu\nu\sigma\rho}T^{\nu\sigma\rho}\ ,\\
t_{\alpha\mu\nu} & \coloneqq  & \frac{1}{2}(T_{\alpha\mu\nu}+T_{\mu\alpha\nu})+\frac{1}{6}(g_{\nu\alpha}v_{\mu}+g_{\nu\mu}v_{\alpha})-\frac{1}{3}g_{\alpha\mu}v_{\nu}\ ,\label{eq:Tor_ir_comp3}
\end{eqnarray}
\end{subequations}
where $\epsilon_{\mu\nu\sigma\rho}$ denotes the Levi-Civita tensor and the tensorial part enjoys the properties  \begin{subequations} \begin{eqnarray}
    t_{\alpha\mu\nu} &=& t_{\mu\alpha\nu}\,, \\
    t_{\alpha\mu\nu} + t_{\nu\alpha\mu} + t_{\mu\nu\alpha} &=& 0\,, \\
    t^{\alpha\mu}_{~~~\alpha} = t^{\alpha~\mu}_{~\alpha} = t^{\mu\alpha}_{~~~\alpha} &=& 0\,.
\end{eqnarray} \end{subequations}
In this irreducible representation we can express the torsion scalar as
\begin{equation}
T = \frac{1}{4}T_1 + \frac{1}{2}T_2 - T_3 = \frac{3}{2}T_{\text{axi}} + \frac{2}{3}T_{\text{ten}} - \frac{2}{3}T_{\text{vec}}\,,
\end{equation}
where 
\begin{subequations}
\begin{eqnarray}\label{eq:Τaxi}
    T_{\text{axi}} & \coloneqq& a_{\mu}a^{\mu}=\frac{1}{18}\left(T_{1}-2T_{2}\right)\,,\\
    T_{\text{vec}} & \coloneqq& v_{\mu}v^{\mu}=T_{3}\,,\\
    T_{\text{ten}} & \coloneqq& t_{\sigma\mu\nu}t^{\sigma\mu\nu}=\frac{1}{2}\left(T_{1}+T_{2}\right)-\frac{1}{2}T_{3}\,.\label{eq:Τten}
\end{eqnarray}
\end{subequations}
Both of these representations are equivalent in general but one is more suitable than the other in specific analyses.

\section{Generalised Proca fields\label{gen_proca_theores}}
\subsection{Generalised Proca in flat space}
In this section, we review generalised Proca theories in standard gravity \cite{Heisenberg:2017mzp,Heisenberg:2014rta,Allys2015}. The goal is to generalise the Proca action in Eq.(\ref{sp}) to include self-interacting terms without changing the propagating degrees of freedom. We will start by defining the Lagrangian for a generalised Proca (GP) vector field as
\begin{equation}
    \mathcal{L}_{GP} = -\frac{1}{4}F^{2} + \sum_{n=2}^{6}\alpha_{n}\mathcal{L}_{n}\,,
\end{equation}
where $F^{2}\equiv F_{\mu\nu}F^{\mu\nu}$, $\alpha_n$ are arbitrary constants and $\mathcal{L}_{n}$ are different self-interacting Lagrangian terms (which will differ in the number of derivatives of the vector field). In order to define the various $\mathcal{L}_{n}$ contributions, one starts with the most general possible combination of vector field and its derivatives, and then reduce their form by imposing physical constraints. In this case, the constraint is given by the fact that we require the generalised theory to propagate 3 degrees of freedom of the the vector field, i.e. there should be no propagation of the zeroth component. To ensure that this is true, the Hessian matrix, a symmetric $n\times n$ matrix of second-order partial derivatives of the scalar field
\begin{equation}
    \mathcal{H}^{\mu\nu}_{\mathcal{L}_{n}} = \frac{\partial^{2}\mathcal{L}_{n}}{\partial\dot{A}_{\mu}\partial\dot{A}_{\nu}}, \ \ \ \text{with} \ \ \ \dot{A}_{\mu} = \partial_{0}A_{\mu}\,,
\end{equation}
is calculated for each $\mathcal{L}_{n}$ and then we impose that the determinant of the matrix vanishes. These conditions ensure that there are no eigenvalues corresponding to the $A_0$ component in the kinetic matrix, thus making the $A_0$ component non-dynamical. The latter statement can also be viewed as requiring $\mathcal{H}^{00} = \mathcal{H}^{0i} = 0$ \cite{Allys2015}. Thus, one finds that the various $\mathcal{L}_{n}$ are given by
\begin{eqnarray}
    \mathcal{L}_{2} &=& f_{2}(X, F, Y)\,,  \label{LGP1}\\
    \mathcal{L}_{3} &=& f_{3}(X)\partial\cdot A\,, \label{LGP2}\\
    \mathcal{L}_{4} &=& f_{4}(X)[(\partial\cdot A)^{2} - \partial_{\mu}A_{\nu}\partial^{\nu}A^{\mu}] + c_{2}\Tilde{f}_{4}(X)F^{2}\,, \label{LGP3}\\
    \mathcal{L}_5 &=& f_{5}(X)[(\partial\cdot A)^{3} - 3(\partial\cdot A)\partial_{\mu}A_{\nu}\partial^{\nu}A^{\mu} +2\partial_{\rho}A_{\sigma}\partial^{\alpha}A^{\rho}\partial^{\sigma}A_{\alpha}] + d_{2}\Tilde{f}_{5}(X)\Tilde{F}^{\mu\rho}\Tilde{F}^{\nu}_{~\rho}\partial_{\mu}A_{\nu}\,, \label{LGP4}\\
    \mathcal{L}_6 &=& e_{2}f_{6}(X)\Tilde{F}^{\mu\nu}\Tilde{F}^{\gamma\rho}\partial_{\mu}A_{\gamma}\partial_{\nu}A_{\rho}\,,\label{LGP5}
\end{eqnarray}
where $\partial\cdot A \equiv \partial_{\mu}A^{\mu}$ will be used hereafter, $X = -A_{\mu}A^{\mu}/2$, $F = F_{\mu\nu}F^{\mu\nu}/4$, $Y = A^{\mu}A^{\nu}F_{\mu}^{\ \alpha}F_{\nu\alpha}$, $f_{2,3,4,5,6}$ are arbitrary functions and $c_{2},\ d_{2},\ e_{2}$ are arbitrary constants. The argument of the arbitrary functions $f_{3,4,5,6}$ is fixed by the fact that these functions are multiplied by derivatives of the vector field, implying that they themselves cannot contain any derivatives of the vector field to avoid any cancellations through partial integration. On the other hand, $f_{2}$ is not multiplied by derivatives of the vector field and thus differs from the other functions. In fact, this function contains all the possible terms which have $U(1)$ symmetry. Also, $X,\ F,\ Y$ are the independent contractions from which all the other terms can be obtained. One should also notice that all of these give rise to second order equations of motion for the vector field, as needed. \medskip

The first Lagrangian term which appears in Eq.(\ref{LGP1}), $\mathcal{L}_{2}$, represents the simplest modification that can be implemented to the action, i.e. promoting the mass term $m$ to a function of the vector field. This term, in fact, includes, among others, the mass term and the potential interactions of the field, $V(A^{2}) \subset f_{2}$. \medskip

The other Lagrangian terms are a broader generalisation of the theory and, in order to understand how these were derived, we will take $\mathcal{L}_{4}$ as an example. We start by considering the most general Lagrangian containing two derivatives of the vector field
\begin{equation}
    \mathcal{L}_{4}=f_{4}(X)[c_{1}(\partial\cdot A)^{2}+c_{2}\partial_{\mu}A_{\nu}\partial^{\mu}A^{\nu}+c_{3}\partial_{\mu}A_{\nu}\partial^{\nu}A^{\mu}]\,,
\end{equation}
where, $c_{1},\ c_{2},\ \text{and} \ c_{3}$ are arbitrary constants. Thus, we find the Hessian matrix
\begin{equation}
\begin{split}
    {H}^{\mu\nu}_{\mathcal{L}_{4}} = f_{4}
    \begin{pmatrix} 2(c_{1} + c_{2} + c_{3}) & 0 \\ 0 & -2c_{2}\delta_{ij}
    \end{pmatrix}\,.
\end{split}
\end{equation}
There exist two possibilities in order to obtain a vanishing determinant. The first is $c_{2} = 0$, but this would imply the propagation of only the temporal component of the vector, which is exactly the opposite of what is required. Thus, we opt for the second option which is the parameter constraint $c_{1} + c_{2} + c_{3} = 0$. Without loss of generality, we can set $c_{1} = 1$, implying $c_{3} = -(1+c_{2})$. Thus, our Lagrangian can then be re-written as
\begin{eqnarray}
    \mathcal{L}_{4} &=& f_{4}[(\partial\cdot A)^{2} + c_{2}\partial_{\mu}A_{\nu}\partial^{\mu}A^{\nu} -(1+ c_{2})\partial_{\mu}A_{\nu}\partial^{\nu}A^{\mu}] \nonumber\\
    &=& f_{4}[(\partial\cdot A)^{2} - \partial_{\mu}A_{\nu}\partial^{\nu}A^{\mu} + c_{2}F^{2}]\,,
\end{eqnarray}
where, the second equality was obtained by simplifying the derivatives. Now, one notices that the term proportional to $c_{2}$ which together with an independent separate function $\Tilde{f}_{4}(X)$ could be included in $f_{2}$. In this case we shall leave the term explicitly in this form, as shown in (\ref{LGP3}). Repeating this process for all the other Langrangian terms, we obtain the Lagrangian setup shown in Eqs.(\ref{LGP1}--\ref{LGP5}).

\subsection{Generalised Proca in curved background} \label{GP_in_curved_spacetime}
The Proca Lagrangian contributions that result in the previous section were obtained on a flat background. We now want to generalise this to a general non-flat background. When going to such a space and making the metric, $g_{\mu\nu}$, dynamical one should also impose that the field equations of the metric have to be, at most, second-order. Furthermore, the derivative self-interactions could lead to possible excitations of the temporal polarisation of the vector field via the formation of non-minimal couplings \cite{Jimenez:2016isa}. Thus, when converting the partial derivatives into covariant ones, one should be careful and add non-minimal couplings to the graviton as counter terms to prevent this, and in order to maintain the second order nature of the equations of motion. \medskip

For the construction of these non-minimal couplings, the divergenceless tensors of the gravity sector are of vital importance. Thus, the total Lagrangian density is given by \cite{Jimenez:2016isa}
\begin{equation}
    \mathcal{L}_{\text{GP}}^{\text{curved}} = -\frac{1}{4}F^{2} + \sum_{n=2}^{6}\beta_{n}\mathcal{L}_{n}\,,
\end{equation}
with \cite{Jimenez:2016isa, Heisenberg:2014rta}
\begin{subequations}
\begin{eqnarray}\label{eq:ProcaL2}
\mathcal{L}_{2} & =&G_{2}(X,F,Y)\,,\\
\mathcal{L}_{3} & =&G_{3}(X)\lc{\nabla}\cdot A\,,\\
\mathcal{L}_{4} & =&G_{4}(X)\lc{R}+ G_{4,X}(X)[(\lc{\nabla}\cdot A)^{2}-\lc{\nabla}_{\mu}A_{\nu}\lc{\nabla}^{\nu}A^{\mu}]\,,\\
\mathcal{L}_{5} & =&G_{5}(X)\lc{G}_{\mu\nu}\lc{\nabla}^{\mu}A^{\nu}-\frac{1}{6}G_{5},_{X}(X)[(\lc{\nabla}\cdot A)^{3}-3(\lc{\nabla}\cdot A)\lc{\nabla}_{\mu}A_{\nu}\lc{\nabla}^{\nu}A^{\mu}\nonumber \\
 && \ \ \ +2\lc{\nabla}_{\mu}A_{\nu}\lc{\nabla}^{\alpha}A^{\mu}\lc{\nabla}^{\nu}A_{\alpha}]-\Tilde{G}_{5}(X)\Tilde{F}^{\alpha\mu}\Tilde{F}_{~\mu}^{\beta}\lc{\nabla}_{\alpha}A_{\beta}\,,\\
\mathcal{L}_{6} & =& G_{6}(X)\lc{L}^{\mu\nu\alpha\beta}\lc{\nabla}_{\mu}A_{\nu}\lc{\nabla}_{\alpha}A_{\beta}+\frac{1}{2}G_{6},_{X}(X)\Tilde{F}^{\alpha\mu}\Tilde{F}^{\beta\gamma}\lc{\nabla}_{\alpha}A_{\beta}\lc{\nabla}_{\mu}A_{\gamma}\,,\label{eq:ProcaL6}
\end{eqnarray}
\end{subequations}
where $f_{,X}\coloneqq \partial f/\partial X$, $\lc{G}_{\mu\nu}$ is the Einstein tensor and $\lc{L}^{\mu\nu\alpha\beta}$ is the double dual Riemann tensor \begin{equation}
    \lc{L}^{\mu\nu\alpha\beta} \coloneqq  \frac{1}{4}\epsilon^{\mu\nu\rho\sigma}\epsilon^{\alpha\beta\gamma\delta}\lc{R}_{\rho\sigma\gamma\delta}\,,
\end{equation}
which inherits the same symmetry properties as the Riemann tensor, i.e. $\lc{L}^{\mu\nu\alpha\beta}=\lc{L}^{\alpha\beta\mu\nu}$, $\lc{L}^{\mu\nu\alpha\beta}=-\lc{L}^{\nu\mu\alpha\beta}$ and $\lc{L}^{\mu\nu\alpha\beta}=-\lc{L}^{\mu\nu\beta\alpha}$ \cite{Allys2015}. Note that the couplings in the last term in $\mathcal{L}_{5}$, $\Tilde{G}_{5}(X)\Tilde{F}^{\alpha\mu}\Tilde{F}_{~\mu}^{\beta}\lc{\nabla}_{\alpha}A_{\beta}$, and $\mathcal{L}_{3}$ do not require any non-minimal counter terms, as in these cases the coupling to the connection is linear \cite{Allys2015, Jimenez:2016isa}. Furthermore, note that terms such as $\lc{G}^{\mu\nu}A_{\mu}A_{\nu}$, which do not propagate the temporal component of the vector field, are already included in the above interactions after integration by parts \cite{Heisenberg:2017mzp}. \medskip

In summary, in this section we have seen how a general theory for Proca fields can be constructed for a flat spacetime and then generalised to a curved one. In the next section, we provide a way to translate this to a TG setting.

\section{Proca theories in teleparallel gravity\label{TG_proca_theory}}

As stated in section \ref{section:Teleparallel Gravity}, the TEGR and GR approach differ in the underlying geometry since the former is a torsion based theory, while the latter is a curvature based one. As seen in section \ref{GP_in_curved_spacetime}, in order to covariantise a theory, one needs to find a connection between the neighbouring tangent spaces to a manifold. Thus, since the geometries are different (implying different connections), the covariantisation procedure differs between the two theories. Nonetheless, since the teleparallel connection is linked to the usual Lorentzian geometry on which GR is based upon via Eq.(\ref{eq:contorsion}), the covariantisation procedure is simplified. \medskip

Keeping as a reference Eq.(\ref{eq:contorsion}), it can be noted that the coupling prescriptions of GR and TG can be taken as equivalent \cite{Aldrovandi:2013wha}. In a similar manner the BDLS theory was formulated \cite{Bahamonde:2019shr,Bahamonde:2019ipm} which we will follow. Since the coupling prescriptions between the teleparallel connection and the Levi-Civita connection are equivalent meaning 
\begin{eqnarray}
    \left.e_{~\mu}^{A}\right|_{flat} && \rightarrow e_{~\mu}^{A}\,,\\
    \partial_{\mu} && \rightarrow\lc{\nabla}_{\mu}\,,
\end{eqnarray}
we use the usual Levi-Civita connection prescription for simplicity. An implication of this fact is that the form of the Eqs.(\ref{eq:ProcaL2}-\ref{eq:ProcaL6}) remains the same under the covariantisation scheme although we will re-write them again with some slight adjustments
\begin{subequations}
\begin{eqnarray}\label{eq:ProcaL2TELE}
\mathcal{L}_{2} & =&G_{2}(X,F,Y)\,,\\
\mathcal{L}_{3} & =&G_{3}(X)\lc{\nabla}\cdot A\,,\\
\mathcal{L}_{4} & =&G_{4}(X)(-T+B)+G_{4,X}(X)[(\lc{\nabla}\cdot A)^{2}-\lc{\nabla}_{\mu}A_{\nu}\lc{\nabla}^{\nu}A^{\mu}]\,,\\
\mathcal{L}_{5} & =&G_{5}(X)\lc{G}_{\mu\nu}\lc{\nabla}^{\mu}A^{\nu}-\frac{1}{6}G_{5},_{X}(X)[(\lc{\nabla}\cdot A)^{3}-3(\lc{\nabla}\cdot A)\lc{\nabla}_{\mu}A_{\nu}\lc{\nabla}^{\nu}A^{\mu}\nonumber \\
 && \ \ \ +2\lc{\nabla}_{\mu}A_{\nu}\lc{\nabla}^{\alpha}A^{\mu}\lc{\nabla}^{\nu}A_{\alpha}]-\Tilde{G}_{5}(X)\Tilde{F}^{\alpha\mu}\Tilde{F}_{~\mu}^{\beta}\lc{\nabla}_{\alpha}A_{\beta}\,,\\
\mathcal{L}_{6} & =& G_{6}(X)\lc{L}^{\mu\nu\alpha\beta}\lc{\nabla}_{\mu}A_{\nu}\lc{\nabla}_{\alpha}A_{\beta}+\frac{1}{2}G_{6},_{X}(X)\Tilde{F}^{\alpha\mu}\Tilde{F}^{\beta\gamma}\lc{\nabla}_{\alpha}A_{\beta}\lc{\nabla}_{\mu}A_{\gamma}\,,\label{eq:ProcaL6TELE}
\end{eqnarray}
\end{subequations}
where we have substituted the Ricci scalar in $\mathcal{L}_4$ from its TG equivalent form as presented in Eq.(\ref{eq:Ricci_torsion_equiv}). One could also use the TG form of the Einstein tensor in the WG as
\begin{equation}
    \lc{G}_{\mu\nu} = e^{-1}e^{a}{}_{\mu}g_{\nu\rho}\partial_{\sigma}(e\dut{S}{a}{\rho\sigma}) - \dudt{S}{b}{\sigma}{\nu} \udt{T}{b}{\sigma\mu} + \frac{1}{4}Tg_{\mu\nu}\,,
\end{equation}
and further express the Riemann tensor in terms of quantities related to TG using the fundamental Eq.(\ref{eq:Ricci_torsion_equiv}) as
\begin{equation}
    \lc{R}{}^{\rho}{}_{\sigma\mu\nu} = -\overset{\circ}{\nabla}{}_{\mu}K{}^{\rho}{}_{\sigma\nu}+\overset{\circ}{\nabla}{}_{\nu}K{}^{\rho}{}_{\sigma\mu}-K^{\beta}{}_{\sigma\nu}K{}^{\rho}{}_{\beta\mu}+K^{\beta}{}_{\sigma\mu}K{}^{\rho}{}_{\beta\nu}\,.
\end{equation}

Note that while these Lagrangians were enough to describe the general Proca theory when using the Levi-Civita connection prescription, the same cannot be said for the teleparallel connection prescription. In fact, when the teleparallel connection is used, the Lovelock's theorem is weakened \cite{Gonzalez:2015sha,Gonzalez:2019tky}, hence dramatically increasing the pool of available scalar invariants constructed via the torsion tensor. These new scalars lead to a new Teleparallel Proca Lagrangian contribution, $\mathcal{L}_{TP}$, hence the new total lagrangian density of the theory $\mathcal{L}_{GP}^{Tele}$ will be given by \begin{equation}
     \mathcal{L}_{\text{GP}}^{\text{Tele}} = -\frac{1}{4}F^{2} + \sum_{n=2}^{6}\beta_{n}\mathcal{L}_{n} + \mathcal{L}_{TP}\ .
\end{equation} 

The form of $\mathcal{L}_{TP}$ can be specified from the scalars we are going to construct out of the irreducible components of the torsion tensor Eq.(\ref{eq:Tor_ir_comp1}-\ref{eq:Tor_ir_comp3}) and the vector field $A_\mu$. Note that our available set of possible scalars is quite large, and thus for the needs of this work we will only focus on scalars built obeying the following rules:
\begin{enumerate}
    \item The resulting field equations must, at most, be of second order in both $e^{A}{}_{\mu}$ and $A_{\mu}$;
    \item $A_{\mu}$ must have maximum 3 degrees of freedom, $A_0$ being not dynamical;
    \item Cannot be parity violating;
    \item Must be linear in the torsion tensor;
    \item Up to fourth order derivatives on $A_{\mu}$, i.e, $\partial A\partial A\partial A\partial A\sim\left(\partial A\right)^{4}$.
\end{enumerate}

The first condition guarantees that we avoid Ostrogradsky ghosts by restricting the Lagrangian to include at most first order covariant derivatives wrt the tetrad and the vector field \cite{Ostrogradsky:1850fid,Woodard:2006nt}. The first condition is not enough since second order equation of motion for the temporal component would mean that $A_{0}$ is a dynamical ghost degree of freedom. Also, it is known that the massive spin-1 representation of the Lorentz group should only carry three dynamical fields and we demand the inclusion of derivative self-interactions does not alter this property. The previous criteria can be combined into the second condition. The fourth and fifth conditions just restrict the number of available scalars that one can include since just including linear in torsion scalars immediately increases the number of available scalars to the order of hundreds. \medskip

\begin{table}[H]
\begin{center}
\setlength{\tabcolsep}{8pt}
\renewcommand{\arraystretch}{1.5}
\begin{tabular}{|c|c|c|c|}
\hline 
$n$ & Vectorial $(v)$ & Axial $(a)$ & Purely tensorial $(t)$\\[2pt]
\hline 
0 & $vA$ & - & $tAAA$\\[2pt]
\hline 
1 & $vAF A$ & $\epsilon aAF A$ & $tAF A$\\[1pt]
\hline 
2 & $vAFF$, $vA\widetilde{F}\widetilde{F}$ & $\epsilon aAFF,\epsilon aA\tilde{F}\tilde{F}$ & $tAFF$,$tA\widetilde{F}\widetilde{F}$\\[2pt]
\hline 
3 & $vAFFF$,$vA\widetilde{F}\widetilde{F}F$ & $\epsilon aAFFF$, $\epsilon aA\tilde{F}\tilde{F}F$ & $tAFFF$,$tA\widetilde{F}\widetilde{F}F$\\[2pt]
\hline 
4 & $vAFFFF$, $vA\widetilde{F}\widetilde{F}\widetilde{F}\widetilde{F}$,
$vA\widetilde{F}\widetilde{F}FF$ & $\epsilon aAFFFF$, $\epsilon aA\tilde{F}\tilde{F}\tilde{F}\tilde{F}$,
$\epsilon aA\tilde{F}\tilde{F}FF$ & $tAFFFF$, $tA\widetilde{F}\widetilde{F}\widetilde{F}\widetilde{F}$,
$tA\widetilde{F}\widetilde{F}FF$\\[2pt]
\hline 
\end{tabular}

\end{center}
\caption{Generators of scalars -- These are the independent components from which all the other terms can be obtained by permuting the indices.\label{table:scalargen}}

\end{table}

Keeping these properties in mind, we gather all the possible contractions schematically in table \ref{table:scalargen} by denoting as $n$ the parameter that indicates the expansion of the product $\underset{n}{\prod}\lc\nabla_{\mu_{n}}A_{\nu_{n}}=\lc\nabla_{\mu_{1}}A_{\nu_{1}}\lc\nabla_{\mu_{2}}A_{\nu_{2}},..,\lc\nabla_{\mu_{n}}A_{\nu_{n}}$ where all possible index configurations are implied for each non-indexed expression. For example the generator $\left\{ vAF\right\}$ spans the following index configurations
\begin{eqnarray}
I_{2} & = & v_{\beta}A_{\alpha}F^{\alpha\beta},\\
I_{3} & = & v_{\beta}A_{\alpha}F^{\beta\alpha},
\end{eqnarray}
one can refer to the appendix \ref{appendix:generators}, for the full expansion of every generator. Note that for $n>1$ the derivatives $\nabla_{\mu} A_{\nu}$ are substituted by $F_{\mu\nu}$ in order to keep the $A_0$ component non-dynamical. We stress that these scalars will only effect non-trivial geometries, away from Minkowski spacetime, where the torsion tensor is not trivialised. Thus, the resulting scalars are trivialised in flat spacetime reducing the theory to the standard generalised Proca one \cite{Heisenberg:2014rta,Heisenberg:2017mzp}. Table \ref{table:scalargen} was partially generated with the help of the xAct packages \cite{Martin-Garcia:2007bqa,MartinGarcia:2008qz,DBLP:journals/corr/abs-0803-0862,Brizuela:2008ra,GomezLobo:2011xv,Pitrou:2013hga,Nutma:2013zea}. \medskip

Having produced the scalars we give the explicit form of $\mathcal{L}_{TP}$, which we will strictly keep linear in any torsion argument, by defining
\begin{equation}\label{TG_Proca_contribution}
    \mathcal{L}_{TP}  \coloneqq G_{TP}\left(X,F,Y,I_1,I_2,..,I_{54}\right)
\end{equation}
where the $I$'s are found in the appendix \ref{appendix:generators} for brevity's sake. Let us point out that this function in a Minkowski setting will be absorbed into $\mathcal L_2$ from Eq.(\ref{eq:ProcaL2TELE}) since all the \textit{I}'s will be trivialized, thus reducing the theory back to usual generalised Proca. At first glance one could argue that there are a lot of $I$ arguments in $G_{TP}\left(X,F,Y,I_1,I_2,..,I_{54}\right)$ but not all of them survive in symmetric backgrounds since most of them are exhaustive permutations of specific forms. This is evident in section \ref{sec:Cosmological_background_in _teleparallel_Proca_Theories} where for a flat cosmological background only 4 of them survive.

\section{Cosmological background in teleparallel Proca Theories}\label{sec:Cosmological_background_in _teleparallel_Proca_Theories}

We now explore the scenario of a flat Friedmann–Lema\^{i}tre–Robertson–Walker (FLRW) cosmology which will expose possible dynamical effects of the vector field at background level. We consider the FLRW metric as $ds^2 = -dt^2 + a(t)^2(dx^2+dy^2+dz^2)$ where $a(t)$ is the scale factor in terms of cosmic time. This can be described by the tetrad
\begin{equation}\label{flrw_tetrad}
    \udt{e}{A}{\mu}=\textrm{diag}(1,a(t),a(t),a(t))\,,
\end{equation}
which is compatible with the so-called WG in which the spin connection components vanish \cite{Krssak:2018ywd}, i.e. $\udt{\omega}{A}{B\mu}=0$. This can be shown to produce the FLRW metric by considering the relations in Eq.(\ref{metric_tetrad_eq}). \medskip

\noindent By taking the definition of the torsion scalar in Eq.(\ref{torsion_scalar_def}), this turns out to be
\begin{equation}
    T = 6H^2\,,
\end{equation}
while the boundary term is given by $B = 6\left(3H^2 + \dot{H}\right)$. This is important because by Eq.(\ref{eq:Ricci_torsion_equiv}), we recover the regular Ricci scalar term as $\lc{R} = -T+B = 6 \left(2H^2 + \dot{H}\right)$. \medskip

\noindent In the present study, we consider the case of a homogeneous and time dependent vector field given by
\begin{equation}
    A_{\mu} = \left(A(t),0,0,0\right)\,,
\end{equation}
where $A(t)$ is a scalar function that depends only on cosmic time. By considering the full breadth of scalars that are produced from table \ref{table:scalargen} (or more specifically by the scalar definitions in appendix (\ref{appendix:generators})), we find that the only nonvanishing scalar is
\begin{eqnarray}
    I_1 &=& 3AH\,,\label{FLRW_scalar1}
\end{eqnarray}
whereas the purely Proca scalars are then given by
\begin{eqnarray}
    X &=& \frac{1}{2}A^2 \,,\\
    Y &=& 0 = F\,.
\end{eqnarray}
This means that the additional Lagrangian contribution will only be effected at background level by these scalars, meaning that $G_{TP} = G_{TP}\left(X,I_1,I_2,I_3,I_4\right)$. \medskip

By considering a Universe filled by a perfect fluid with energy density $\rho$ and pressure $p$. The ensuing Friedmann equations turn out to be
\begin{eqnarray}
    \mathcal{A}_{\rm TP} + \Sigma_{i=2}^5 \mathcal{A}_i &=& \rho\,,\label{Friedmann1}\\
    \mathcal{B}_{\rm TP} + \Sigma_{i=2}^5 \mathcal{B}_i &=& p\,,\label{Friedmann2}
\end{eqnarray}
where
\begin{eqnarray}
\mathcal{A}_{2} & = & G_{2}-A^{2}G_{2,{\rm {X}}}\,,\\
\mathcal{A}_{3} & = & -3HA^{3}G_{3,{X}}\,,\\
\mathcal{A}_{4} & = & 6H^{2}G_{4}-6\left(2G_{4,{X}}+G_{4,{XX}}A^{2}\right)H^{2}A^{2}\,,\\
\mathcal{A}_{5} & = & G_{5,{XX}}H^{3}A^{5}+5G_{5,{X}}H^{3}A^{3}\,,\\
\mathcal{A}_{{\rm TP}} & = & G_{{TP}}-A\left(AG_{{\rm {TP,X}}}+6HG_{{\rm {TP,I_{1}}}}\right)\,,
\end{eqnarray}
for the first Friedmann equation, where commas denote derivatives with respect to the appearing scalars, and 
\begin{eqnarray}
\mathcal{B}_{2} & = & G_{2}\,,\\
\mathcal{B}_{3} & = & -A\dot{A}G_{3,{\rm X}}\,,\\
\mathcal{B}_{4} & = & 2G_{4}\left(3H^{2}+2\dot{H}\right)-2G_{4,{\rm X}}\left(3H^{2}A+2H\dot{A}+2\dot{H}A\right)-4G_{4,{\rm XX}}HA^{3}\dot{A}\,,\\
\mathcal{B}_{5} & = & G_{5,{\rm XX}}H^{2}A^{4}\dot{A}+G_{5,{\rm X}}HA^{2}\left(2\dot{H}A+2H^{2}A+3H\dot{A}\right)\,,\\
\mathcal{B}_{{\rm TP}} & = & G_{{TP}}-3AG_{{\rm {TP,I_{1}I_{1}}}}\left(H\dot{A}+A\dot{H}\right)-G_{{\rm {TP,I_{1}}}}\left(\dot{A}+3AH\right)-A{}^{2}\dot{A}G_{{\rm {TP,I_{1}X}}}\,,
\end{eqnarray}
for the second Friedmann equation. Similarly, the vector conservation equation can be obtained, giving
\begin{equation}
    \mathcal{P}_{\rm TP} + \Sigma_{i=2}^5 \mathcal{P}_i = 0\,,\label{vector_conservation_eq}
\end{equation}
where
\begin{eqnarray}
\mathcal{P}_{2} & = & AG_{2,{\rm X}}\,,\\
\mathcal{P}_{3} & = & 3A^{2}HG_{3,{\rm X}}\,,\\
\mathcal{P}_{4} & = & 6AH^{2}G_{4,{\rm X}}+6H^{2}A^{3}G_{4,{\rm XX}}\,,\\
\mathcal{P}_{5} & = & -3H^{3}A^{2}G_{5,{\rm X}}-H^{3}A^{4}G_{5,{\rm XX}}\,,\\
\mathcal{P}_{{\rm TP}} & = & AG_{{\rm {TP,X}}}+3HG_{{\rm {TP,I_{1}}}}\,.
\end{eqnarray}
Altogether, Eqs.(\ref{Friedmann1},\ref{Friedmann2},\ref{vector_conservation_eq}) represent the background evolution equations that describe the teleparallel Proca theory being proposed in this work. In the limit where the new terms vanish, this tends to the standard gravity version of Proca gravity \cite{Heisenberg:2014rta}, as expected. Despite the low number of nonvanishing scalars, the resulting cosmology immediately becomes very rich even at background level indicating even more possible impacts of these new scalars at perturbative order.

\section{Conclusion}\label{Conclusion}
Proca theories of gravity offer an interesting framework in which to produce gravitational theories that includes a massive spin-1 vector field. In terms of second order equations of motion, this lays out the Lagrangian contributions in Eqs.(\ref{LGP1}--\ref{LGP5}), which together describe the generalised Proca theory on Minkowski spacetime. By using the minimal coupling prescription, this can be directly extended to an arbitrary curvature spacetime that results in Eqs.(\ref{eq:ProcaL2}--\ref{eq:ProcaL6}). These equations describe the covariantised version of the generalised Proca theory which is made up of scalars constructed from the vector field and coupling functions. \medskip

The TG analog of the standard gravity version of Proca theories is obtained by using the TG coupling prescription to raise the Minkowski spacetime contributions in Eqs.(\ref{LGP1}--\ref{LGP5}) which would produce an infinite number of terms due to the lower order nature of torsional theories of gravity. In this context, we introduce the five guiding principles on which we build our theory, namely that (i) second order equations of motion are produced, in order to avoid Ostrogradsky instabilities; (ii) only three degrees of freedom are expressed via the vector field, which is a core requirement of Proca theories; (iii) does not violate parity, which is a good base for healthy theories; (iv) Lagrangian contributions must be, at most, linear in the torsion tensor, since higher order contractions may not contribute to leading order terms; (v) vector field derivatives must appear less than fourth order so that it is related to the structure of the generalised Proca theory thus providing a reasonable cutoff. Altogether these conditions produce a well defined TG generalised Proca theory where the standard gravity terms are complemented by an additional contribution in the form of the Lagrangian term in Eq.(\ref{TG_Proca_contribution}). TG is very sensitive to the selection rules that forms the scalars since even these seemingly selective rules produce a large number of potential scalars. \medskip

One may loosen the selection criteria and produce arbitrarily large additional contributions to the standard  Proca action which may take the form of further contributions of the vector coupling terms. Another possibility is the inclusion of quadractic(or even arbitrary higher) TG scalars as arguments, like the irreducible torsion scalars Eqs.(\ref{eq:Τaxi}--\ref{eq:Τten})  \cite{Hayashi:1979qx,Blixt:2019ene}. In this context, the present new Lagrangian term in Eq.(\ref{TG_Proca_contribution}) is simply one of the myriad of potential expressions of generalised Proca theories in TG which offers a large class of potential avenues for producing viable models that may produce a fundamental field theory explanation for a number of important phenomena such as dark energy and dark matter. \medskip

Finally, in section \ref{sec:Cosmological_background_in _teleparallel_Proca_Theories} we present a simple example of a homogeneous and isotropic Universe in which we derive the Friedmann equations. These background level equations are important for determining the expansion rate of the Universe, but also to assess the sensitivity of the equations of motion to the various scalars being proposed in this work. We find that only four scalars contribute in a nonvanishing way to the equations of motion, which are presented in Eqs.(\ref{FLRW_scalar1}--\ref{FLRW_scalar2}) where a vector field that changes only in time is assumed. This means that the vector field scalars only produce a nonvanishing $X$ contribution. Despite this overly simplified scenario, the Friedmann equations still turn out to be quite involved. The first and second Friedmann equations are presented in Eq.(\ref{Friedmann1}--\ref{Friedmann2}) whereas the vector equation in given in Eq.(\ref{vector_conservation_eq}). The combination of these three equations fully determines the cosmology at background level. \medskip

It would be interesting probe this new class of generalised Proca theories and investigate how the models from standard gravity are extended in this framework. This will have ramifications beyond background level and may drastically change the cosmological perturbations for such theories.

\subsection*{Acknowledgements}
JLS would also like to acknowledge funding support from Cosmology@MALTA which is supported by the University of Malta. The authors would like to acknowledge networking support by the COST Action CA18108. V.G would like to thank J. Beltran for useful and fruitful discussions.

\newpage

\appendix

\section{Teleparallel Proca scalars}\label{appendix:generators}

\noindent In this appendix we will expand all the generators from the table \ref{table:scalargen} in all their possible index configurations. We will denote the generator or groups of generators with brackets like in the example after the table \ref{table:scalargen}, $\left\{ vA\lc\nabla A\right\}$. Note that the following sets of scalars are the full list of possible independent scalars. 

\subsection{\textmd{\normalsize{}Torsion vector component \texorpdfstring{$v_{\mu}$}{}}}

\begin{eqnarray}
 &  & \left\{ vA\right\} \nonumber \\
 &  & I_{1}\coloneqq v_{\mu}A^{\mu},\\
\nonumber \\
 &  & \left\{ vAF\right\} \nonumber \\
 &  & I_{2}\coloneqq A_{\alpha}v{}_{\beta}F^{\alpha\beta},\\
 &  & I_{3}\coloneqq A_{\alpha}v{}_{\beta}F^{\beta\alpha},\\
\nonumber \\
 &  & \left\{ vAFF,vA\widetilde{F}\widetilde{F}\right\} \nonumber \\
 &  & I_{4}\coloneqq A_{\alpha}F{}^{2}v{}^{\alpha},\\
 &  & I_{5}\coloneqq A_{\alpha}F{}^{\alpha}{}_{\gamma}F{}^{\beta\gamma}v{}_{\beta},\\
\nonumber \\
 &  & \left\{ vAFFF,vA\widetilde{F}\widetilde{F}F\right\} \nonumber \\
 &  & I_{6}\coloneqq A_{\alpha}F{}^{\alpha}{}_{\mu}F{}^{\beta}{}_{\gamma}F{}^{\mu\gamma}v{}_{\beta},\\
 &  & I_{7}\coloneqq A_{\alpha}F{}^{\alpha\beta}F{}^{2}v{}_{\beta},\\
\nonumber \\
 &  & \left\{ vAFFFF,vA\widetilde{F}\widetilde{F}\widetilde{F}\widetilde{F},vA\widetilde{F}\widetilde{F}FF\right\} \nonumber \\
 &  & I_{8}\coloneqq A_{\alpha}F{}_{\beta\nu}F{}_{\gamma}{}^{\beta}F{}^{\gamma\mu}F{}_{\mu}{}^{\nu}v{}^{\alpha},\\
 &  & I_{9}\coloneqq A_{\alpha}F{}^{4}v{}^{\alpha},\\
 &  & I_{10}\coloneqq A_{\alpha}F{}^{\alpha}{}_{\mu}F{}^{\beta}{}_{\gamma}F{}^{\gamma\nu}F{}^{\mu}{}_{\nu}v{}_{\beta},\\
 &  & I_{11}\coloneqq A_{\alpha}F{}^{\alpha}{}_{\gamma}F{}^{\beta\gamma}F{}^{2}v{}_{\beta},
\end{eqnarray}

\subsection{\textmd{\normalsize{}Torsion axial component \texorpdfstring{$a_{\mu}$}{}}}

\begin{eqnarray}
 &  & \left\{ \epsilon aAF\right\} \nonumber \\
 &  & I_{12}\coloneqq A^{\alpha}a{}^{\beta}\epsilon{}_{\alpha\beta\gamma\mu}F^{\mu\gamma},\\
\nonumber \\
 &  & \left\{ \epsilon aAFF,\epsilon aA\tilde{F}\tilde{F}\right\} \nonumber \\
 &  & I_{13}\coloneqq A_{\alpha}a{}_{\beta}\epsilon{}^{\beta\mu\nu\gamma}F{}^{\alpha}{}_{\mu}F{}_{\nu\gamma},\\
 &  & I_{14}\coloneqq A_{\alpha}a{}_{\beta}\epsilon{}^{\alpha\mu\nu\gamma}F{}^{\beta}{}_{\mu}F{}_{\nu\gamma},\\
 &  & I_{15}\coloneqq A_{\alpha}a{}^{\alpha}\epsilon{}^{\gamma\mu\nu\beta}F{}_{\gamma\mu}F{}_{\nu\beta},\\
\nonumber \\
 &  & \left\{ \epsilon aAFFF,\epsilon aA\tilde{F}\tilde{F}F\right\} \nonumber \\
 &  & I_{16}\coloneqq A_{\alpha}a{}_{\beta}\epsilon{}^{\alpha\beta\mu\nu}F{}_{\gamma\rho}F{}^{\gamma}{}_{\mu}F{}^{\rho}{}_{\nu},\\
 &  & I_{17}\coloneqq A_{\alpha}a{}_{\beta}\epsilon{}^{\mu\rho\gamma\nu}F{}^{\alpha}{}_{\mu}F{}^{\beta}{}_{\rho}F{}_{\gamma\nu},\\
 &  & I_{18}\coloneqq A_{\alpha}a{}_{\beta}\epsilon{}^{\beta\rho\gamma\nu}F{}^{\alpha}{}_{\mu}F{}_{\gamma\nu}F{}^{\mu}{}_{\rho},\\
 &  & I_{19}\coloneqq A_{\alpha}a{}_{\beta}\epsilon{}^{\alpha\rho\gamma\nu}F{}^{\beta}{}_{\mu}F{}_{\gamma\nu}F{}^{\mu}{}_{\rho},\\
 &  & I_{20}\coloneqq A_{\alpha}a{}_{\beta}\epsilon{}^{\nu\rho\gamma\mu}F{}^{\alpha\beta}F{}_{\gamma\mu}F{}_{\nu\rho},\\
 &  & I_{21}\coloneqq A_{\alpha}a{}_{\beta}\epsilon{}^{\alpha\beta\nu\rho}F{}_{\nu\rho}F{}^{2},\\
\nonumber \\
 &  & \left\{ \epsilon aAFFFF,\epsilon aA\tilde{F}\tilde{F}\tilde{F}\tilde{F},\epsilon aA\tilde{F}\tilde{F}FF\right\} \nonumber \\
 &  & I_{22}\coloneqq A_{\alpha}a{}_{\beta}\epsilon{}^{\beta\mu\rho\sigma}F{}^{\alpha}{}_{\mu}F{}^{\gamma}{}_{\sigma}F{}_{\nu\gamma}F{}^{\nu}{}_{\rho},\\
 &  & I_{23}\coloneqq A_{\alpha}a{}_{\beta}\epsilon{}^{\alpha\mu\rho\sigma}F{}^{\beta}{}_{\mu}F{}^{\gamma}{}_{\sigma}F{}_{\nu\gamma}F{}^{\nu}{}_{\rho},\\
 &  & I_{24}\coloneqq A_{\alpha}a{}_{\beta}\epsilon{}^{\rho\gamma\nu\sigma}F{}^{\alpha}{}_{\mu}F{}^{\beta}{}_{\rho}F{}^{\mu}{}_{\gamma}F{}_{\nu\sigma},\\
 &  & I_{25}\coloneqq A_{\alpha}a{}^{\alpha}\epsilon{}^{\mu\beta\nu\sigma}F{}_{\gamma\rho}F{}^{\gamma}{}_{\mu}F{}_{\nu\sigma}F{}^{\rho}{}_{\beta},\\
 &  & I_{26}\coloneqq A_{\alpha}a{}_{\beta}\epsilon{}^{\mu\gamma\nu\sigma}F{}^{\alpha}{}_{\mu}F{}^{\beta}{}_{\rho}F{}_{\nu\sigma}F{}^{\rho}{}_{\gamma},\\
 &  & I_{27}\coloneqq A_{\alpha}a{}_{\beta}\epsilon{}^{\beta\gamma\nu\sigma}F{}^{\alpha}{}_{\mu}F{}^{\mu}{}_{\rho}F{}_{\nu\sigma}F{}^{\rho}{}_{\gamma},\\
 &  & I_{28}\coloneqq A_{\alpha}a{}_{\beta}\epsilon{}^{\alpha\gamma\nu\sigma}F{}^{\beta}{}_{\mu}F{}^{\mu}{}_{\rho}F{}_{\nu\sigma}F{}^{\rho}{}_{\gamma},\\
 &  & I_{29}\coloneqq A_{\alpha}a{}_{\beta}\epsilon{}^{\sigma\gamma\nu\rho}F{}^{\alpha}{}_{\mu}F{}^{\beta\mu}F{}_{\nu\rho}F{}_{\sigma\gamma},\\
 &  & I_{30}\coloneqq A_{\alpha}a{}^{\alpha}\epsilon{}^{\nu\rho\sigma\beta}F{}_{\nu\rho}F{}_{\sigma\beta}F{}^{2},\\
 &  & I_{31}\coloneqq A_{\alpha}a{}_{\beta}\epsilon{}^{\beta\rho\sigma\gamma}F{}^{\alpha}{}_{\rho}F{}_{\sigma\gamma}F{}^{2},\\
 &  & I_{32}\coloneqq A_{\alpha}a{}_{\beta}\epsilon{}^{\alpha\rho\sigma\gamma}F{}^{\beta}{}_{\rho}F{}_{\sigma\gamma}F{}^{2},
\end{eqnarray}

\subsection{\textmd{\normalsize{}Purely tensorial component \texorpdfstring{$t_{\alpha\beta\gamma}$}{}}}

\begin{eqnarray}
 &  & \left\{ tAAA\right\} \nonumber \\
 &  & I_{33}\coloneqq A^{\alpha}A{}^{\beta}A{}^{\gamma}t{}_{\alpha\beta\gamma},\\
\nonumber \\
 &  & \left\{ tAF\right\} \nonumber \\
 &  & I_{34}\coloneqq A^{\alpha}t{}_{\alpha\beta\gamma}F^{\gamma\beta},\\
 &  & I_{35}\coloneqq A^{\alpha}t{}_{\alpha\gamma\beta}F^{\gamma\beta},\\
 &  & I_{36}\coloneqq A^{\alpha}t{}_{\beta\gamma\alpha}F^{\gamma\beta},\\
\nonumber \\
 &  & \left\{ tAFF,tA\widetilde{F}\widetilde{F}\right\} \nonumber \\
 &  & I_{37}\coloneqq A_{\alpha}F{}_{\beta\mu}F{}^{\beta}{}_{\gamma}t{}^{\alpha\mu\gamma},\\
 &  & I_{38}\coloneqq A_{\alpha}F{}^{\alpha}{}_{\beta}F{}_{\gamma\mu}t{}^{\beta\gamma\mu},\\
 &  & I_{39}\coloneqq A_{\alpha}F{}_{\beta\mu}F{}^{\beta}{}_{\gamma}t{}^{\gamma\mu\alpha},\\
\nonumber \\
 &  & \left\{ tAFFF,tA\widetilde{F}\widetilde{F}F\right\} \nonumber \\
 &  & I_{40}\coloneqq A_{\alpha}F{}_{\beta\mu}F{}^{\beta}{}_{\gamma}F{}^{\mu}{}_{\nu}t{}^{\alpha\gamma\nu},\\
 &  & I_{41}\coloneqq A_{\alpha}F{}_{\beta\gamma}F{}^{2}t{}^{\alpha\beta\gamma},\\
 &  & I_{42}\coloneqq A_{\alpha}F{}^{\alpha}{}_{\beta}F{}_{\gamma\nu}F{}^{\gamma}{}_{\mu}t{}^{\beta\nu\mu},\\
 &  & I_{43}\coloneqq A_{\alpha}F{}^{\alpha}{}_{\beta}F{}^{\beta}{}_{\gamma}F{}_{\mu\nu}t{}^{\gamma\mu\nu},\\
 &  & I_{44}\coloneqq A_{\alpha}F{}^{\alpha}{}_{\beta}F{}_{\gamma\nu}F{}^{\gamma}{}_{\mu}t{}^{\mu\nu\beta},\\
\nonumber \\
 &  & \left\{ tAFFFF,tA\widetilde{F}\widetilde{F}\widetilde{F}\widetilde{F},tA\widetilde{F}\widetilde{F}FF,tAFFFF\right\} \nonumber \\
 &  & I_{45}\coloneqq A_{\alpha}F{}_{\beta}{}^{\mu}F{}^{\beta}{}_{\gamma}F{}_{\mu\nu}F{}^{\nu}{}_{\rho}t{}^{\alpha\rho\gamma},\\
 &  & I_{46}\coloneqq A_{\alpha}F{}_{\beta\mu}F{}^{\beta}{}_{\gamma}F{}^{2}t{}^{\alpha\mu\gamma},\\
 &  & I_{47}\coloneqq A_{\alpha}F{}^{\alpha}{}_{\beta}F{}_{\gamma\nu}F{}^{\gamma}{}_{\mu}F{}^{\nu}{}_{\rho}t{}^{\beta\mu\rho},\\
 &  & I_{48}\coloneqq A_{\alpha}F{}^{\alpha}{}_{\beta}F{}_{\gamma\mu}F{}^{2}t{}^{\beta\gamma\mu},\\
 &  & I_{49}\coloneqq A_{\alpha}F{}^{\alpha}{}_{\beta}F{}^{\beta}{}_{\gamma}F{}_{\mu\rho}F{}^{\mu}{}_{\nu}t{}^{\gamma\rho\nu},\\
 &  & I_{50}\coloneqq A_{\alpha}F{}^{\alpha}{}_{\beta}F{}^{\beta}{}_{\gamma}F{}^{\gamma}{}_{\mu}F{}_{\nu\rho}t{}^{\mu\nu\rho},\\
 &  & I_{51}\coloneqq A_{\alpha}F{}_{\beta}{}^{\mu}F{}^{\beta}{}_{\gamma}F{}_{\mu\nu}F{}^{\nu}{}_{\rho}t{}^{\gamma\rho\alpha},\\
 &  & I_{52}\coloneqq A_{\alpha}F{}_{\beta\mu}F{}^{\beta}{}_{\gamma}F{}^{2}t{}^{\gamma\mu\alpha},\\
 &  & I_{53}\coloneqq A_{\alpha}F{}^{\alpha}{}_{\beta}F{}^{\beta}{}_{\gamma}F{}_{\mu\rho}F{}^{\mu}{}_{\nu}t{}^{\nu\rho\gamma}.
\end{eqnarray}

\bibliographystyle{utphys}
\providecommand{\href}[2]{#2}\begingroup\raggedright\endgroup

\end{document}